\documentclass[]{article}

\usepackage{arxiv}

\usepackage[utf8]{inputenc} 
\usepackage[T1]{fontenc}    
\usepackage{hyperref}       
\usepackage{url}            
\usepackage{amsfonts}       
\usepackage{microtype}      
\usepackage{lipsum}		
\usepackage{graphicx}
\usepackage{natbib}
\usepackage{doi}

\begin{document}

\title{Risks of AI Foundation Models in Education}

\author{
    Su Lin Blodgett \\
	Microsoft Research\\
	Montr\'{e}al, Canada\\
	\And
    Michael Madaio\\
	Microsoft Research\\
	New York, NY \\
}

\date{}

\renewcommand{\shorttitle}{Risks of AI Foundation Models in Education}

\maketitle

If the authors of a recent Stanford report \citep{bommasani2021opportunities} on the opportunities and risks of ``foundation models'' are to be believed, these models represent a paradigm shift for AI and for the domains in which they will supposedly be used, including education. Although the name is new (and contested \citep{field2021at}), the term describes existing types of algorithmic models that are ``trained on broad data at scale''and ``fine-tuned'' (i.e., adapted) for particular downstream tasks, and is intended to encompass large language models such as BERT or GPT-3 and computer vision models such as CLIP. Such technologies have the potential for harm broadly speaking \citep[e.g.,][]{bender2021dangers}, but their use in the educational domain is particularly fraught, despite the potential benefits for learners claimed by the authors. In section 3.3 of the Stanford report, Malik et al. argue that achieving the goal of providing education for all learners requires more efficient computational approaches that can rapidly scale across educational domains and across educational contexts, for which they argue foundation models are uniquely well-suited. However, evidence suggests that not only are foundation models not likely to achieve the stated benefits for learners, but their use may also introduce new risks for harm.

That is, even if foundation models work as described, the nature of their design and use may lead to a homogenization of learning in ways that perpetuate the inequitable status quo in education \citep[cf.][]{madaio2021structural}, and which may lead to increasingly limited opportunities for meaningful roles for educational stakeholders in their design. In addition, the all-encompassing vision for foundation models in education laid out by Malik et al. may privilege those aspects of education that are legible to large-scale data collection and modeling, further motivating increased surveillance of children under the guise of care and further devaluing the human experiences of learning that cannot be ingested into foundation models. 

\section{Risks of educational technologies at scale}

While perhaps well-meaning, the argument that developing large-scale models for education is for students' benefit is an argument that is often used to motivate the increasing (and increasingly harmful) use of educational technologies, particularly educational surveillance technologies \citep[cf.][]{collins2021privacy}. Recent research suggests that equity is used by educational AI researchers as a motivation for designing and deploying educational AI, but it is rarely studied as a downstream effect of such technologies \citep{holmes2021ethics}. Indeed, educational AI, including computer vision and natural language technologies, are increasingly deployed to monitor students in the classroom \citep{galligan2020cameras}, at their homes during high-stakes assessments \citep{cahn2020snooping,swauger2020our,barrett2021rejecting}, and to monitor their digital communication across platforms \citep{keierleber2021exclusive}. These applications are often motivated by appeals to care for students' well-being \citep{collins2021privacy}, academic integrity \citep{harwell2020cheating}, or supporting teachers \citep{ogan2019reframing}; and yet, these noble goals do not prevent these technologies from causing harm to learners. As we argue in a recent paper \citep{madaio2021structural}, the fundamental assumptions of many educational AI systems, while motivated by logics of care, may instead reproduce structural inequities of the status quo in education.

In addition to students' well-being, the argument for developing foundation models for education relies on economic logics of efficiency --- that given the rising costs of education, approaches are needed that can provide education ``at scale.'' However, this argument has been made many times before from educational technologists and education reformers, with outcomes that have not, in fact, benefited all learners. It is thus worth taking seriously the lessons of historical arguments for using technology to provide education at scale. Historians of educational technology such as Larry Cuban and Audrey Watters have argued that a century of educational technologies supposedly designed to provide more efficient learning (e.g., educational radio, TV, and computers) have instead led to widespread second-order effects, such as promoting more impersonal, dehumanizing learning experiences and further de-valuing the role of teachers, counterbalanced by teachers' widespread resistance to the adoption and use of these technologies \citep{cuban1986teachers,watters2021teaching}.

Indeed, one can look to recent claims about Massive Open Online Courses (MOOCs) to provide education for learners around the world who could not afford to access a university \citep{pappano2012year}. Although this may have been true for some learners, the vast majority of learners who used and completed MOOCs were already well-educated learners from the Global North, and the majority of courses on major MOOC platforms were offered in English by North American institutions \citep{kizilcec2017closing}. Even for learners who had access to them, the design and use of MOOC platforms thus amplified particular ideologies about teaching and learning, including an ``instructionist'' teaching paradigm with the use of lecture videos that learning science research suggests is less effective than active learning \citep{koedinger2015learning} and which may not be effective across multiple cultural contexts. More generally, other (non-digital) technologies of scale, such as educational standards (e.g., Common Core State Standards) and standardized testing such as the SATs act as ``racializing assemblages'' \citep{dixon2019racializing} that may reproduce sociopolitical categories of difference in ways that reinforce longstanding social hierarchies.

These histories teach us that we should critically interrogate claims about the ability for technology to provide education at scale for all learners, much like claims about the benefits of technologies of scale more generally \citep{hanna2020against}. In addition to interrogating the potential harms of the scalar logics of foundation models, several of which we identify here, we also suggest interrogating who benefits from this drive to scale, and what alternatives it forecloses. Devoting time and resources towards educational technologies built atop foundation models not only diverts our attention away from other educational technologies we might develop (or the question of whether we should develop educational technology at all), but further entrenches the status quo, allowing us to avoid asking hard questions about how existing educational paradigms shape learning processes and outcomes in profoundly inequitable ways \citep{madaio2021structural}. 

\section{Risks of homogenization}

The adaptability of foundation models, where a few large-scale pre-trained models enable a wide range of applications, brings with it particular risks of homogenization\footnote{Here we use ``homogenization'' to refer to the homogenization of outcomes emerging from the use of foundation models, as used in Section 5 of the report (as opposed to the homogenization of models and approaches, as used in Section 1).} for educational applications. That is, design decisions for foundation models --- including decisions about tasks, training data, model output, and more --- may lead to homogenization of pedagogical approaches, of ideologies about learning, and of educational content in ways that may perpetuate existing inequities in education \citep[cf.][]{madaio2021structural}, particularly when such technologies are intended to be deployed ``at scale'' across contexts.  

Specifically, the choices of data proposed for pre-trained models for education smuggle in particular ideologies of teaching and learning that may promote a homogenized vision of instruction in similar ways as previous technologies of education at scale reproduced instructionist modes of teaching\footnote{Indeed, Malik et al., propose using lecture videos from online courses to train instructional models.} \citep{koedinger2015learning}, which may lead to downstream harms for learners, despite claims for these technologies' benefits. For example, Malik et al. propose using feedback provided to developers on forums such as StackOverflow to train feedback models, but the feedback provided on such forums may not be delivered in pedagogically effective ways, and often reproduces toxic cultures of masculinity in computer science in ways that actively exclude novice developers and women \citep{ford2016paradise}. 

Finally, much as \citet{bender2021dangers} have observed for NLP training corpora more generally, the corpora suggested by Malik et al. for training foundation models, such as Project Gutenberg, include texts largely written in English \citep{gerlach2020standardized}, which may produce representational harms \citep[cf.][]{blodgett-etal-2020-language} by reproducing dominant perspectives and language varieties and excluding others. Similarly, \citet{dodge2021documenting} have found that a filter used to create the Colossal Clean Crawled Corpus (C4, a large web-crawled corpus used to train large English language models), ``disproportionately removes documents in dialects of English associated with minority identities (e.g., text in African American English, text discussing LGBTQ+ identities).'' In reproducing socially dominant language varieties, foundation models may require speakers of minoritized varieties to accommodate to dominant varieties in educational contexts, incurring higher costs for these speakers and denying them the legitimacy and use of their varieties \citep{baker-bell2020linguistic}. One setting in which such harms are likely to arise is the use of foundation models for feedback on open-ended writing tasks, as the authors of the report propose. In other work in this space, automated essay scoring and feedback provision have been shown to have roots in racialized histories of writing assessments that are difficult for data-driven technologies trained on such rubrics to avoid \citep{dixon2019racializing}, and automated approaches to writing scoring and feedback may induce students to adopt writing styles that mirror dominant cultures \citep{mayfield2019individual}. 

In this way, foundation models may reproduce harmful ideologies about what is valuable for students to know and how students should be taught, including ideologies about the legitimacy and appropriateness of minoritized language varieties. Given the broader risks of homogenization of foundation models, they may amplify these ideologies at scale.  

\section{Risks of limited roles of stakeholders in designing foundation models}

In education, decisions about educational curricula and pedagogy are often made with sustained critical evaluation and public debates about what to teach and how to teach it \citep[cf.][]{scribner2016fight}. However, by relying on foundation models for broad swaths of education (as Malik et al. propose), decisions about what is to be taught and how students should be taught may be made without the involvement of teachers or other educational stakeholders. Despite claims elsewhere in the Stanford report for foundation models to support human-in-the-loop paradigms [section 2.4], the pre-trained paradigm of foundation models will likely entail limited opportunities for educational stakeholders to participate in key upstream design decisions, limiting their involvement to the use of models once such models are trained or fine-tuned. As with AI more generally, despite rhetoric about the importance of stakeholder participation in designing AI systems \citep{kulynych2020participatory}, the reality of current industrial paradigms of training foundation models on massive datasets requiring massive (and expensive) compute power may limit stakeholders' ability to meaningfully shape choices about tasks, datasets, and model evaluation. 

This narrow scope for involvement of key stakeholders such as teachers and students is at odds with participatory, learner-centered paradigms from educational philosophy \citep[e.g.,][]{freire1996pedagogy,broughan2020re} and the learning sciences \citep{disalvo2017participatory}, where learners' interests and needs shape teachers' choices about what and how to teach. In addition, this may have the effect of further disempowering teachers from having meaningful agency over choices about content or pedagogy, further contributing to the deskilling of the teaching profession, in ways seen in earlier technologies of scale in education \citep[cf.][]{cuban1986teachers,watters2021teaching}.

\section{Risks of totalizing visions of foundation models in education}

All of this raises concerns about the expansive claims made for the application of foundation models in education to ``understand'' students, educators, learning, teaching, and ``subject matter.'' The list of potential uses of foundation models in education claimed by Malik et al. is evocative of the totalizing rhetoric popular in computer science more generally, such as for software to ``eat the world'' \citep{andreessen2011why}. Crucially, this totalizing vision suggests that everything that matters to learning can be rendered into grist for foundation models.  

First, we note that in the education section, ``understanding'' is used to refer to at least three distinct phenomena: students' ``understanding'' of subject matter, foundation models’ ``understanding'' of subject matter, and foundation models' ``understanding'' of pedagogy and student behavior. But as \citet{bender-koller-2020-climbing} have argued, NLP systems (including foundation models) do not have the capability to ``understand'' language or human behavior. Although the pattern matching that underlies foundation models (for NLP or otherwise) may produce outputs that resemble human understanding, this rhetorical slippage is particularly harmful in educational contexts. Because supporting learners' conceptual understanding is one primary goal of education, the conflation of models' representational ability with comprehension of subject matter, as well as the conflation of students' development of conceptual or procedural knowledge with foundation models' pattern-matching capabilities, may lead teachers and other educational stakeholders to trust the capabilities of foundation models in education when such trust is not warranted. 

More generally, the paradigm of foundation models as laid out by Malik et al. requires that teaching and learning be formalized in ways that are legible to foundation models, without interrogating the potential risks of formalizing teaching and learning in such a way, nor the risks for fundamental aspects of education to be discarded if they do not fit into this paradigm. In other words, what forms of knowledge are privileged by rendering them tractable to foundation models? What is lost in such a partial, reductive vision of teaching and learning?

As one example, foundation models may be able to reproduce patterns of pedagogical decisions in the training corpora, but those datasets or models may not be able to capture why those decisions were made. For instance, good teachers draw on a wealth of contextual information about their students' lives, motivations, and interests; information which may not be legible to foundation models. In some cases, the response from AI researchers may be to simply collect more data traces on students in order to make these aspects of students' lives legible to modeling; however, this ``rapacious''\footnote{\url{https://mobile.twitter.com/hypervisible/status/1442473891381710858}} approach to data collection is likely to harm students through the ever-increasing surveillance of students \citep{galligan2020cameras,barrett2021rejecting}.

Despite expansive claims for the potential for foundation models to radically transform teaching and learning in ways that benefit learners, the history of educational technologies suggests that we should approach such claims with a critical eye. In part, the proposed application of foundation models for education brings with it risks for reproducing and amplifying the existing inequitable status quo in education, as well as risks of reproducing dominant cultural ideologies about teaching and learning, in ways that may be harmful for minoritized learners. In addition, the properties of the foundation model paradigm that lend it its appeal --- large-scale, pre-trained models adaptable to downstream tasks --- are precisely what would likely limit opportunities for meaningful participation of teachers, students, and other education stakeholders in key decisions about their design. Education is a fundamentally public good; rather than centralizing power in the hands of institutions with sufficient resources to develop large-scale models, educational technologies, if they are to be designed, should be designed in ways that afford more public participation and are responsive to the needs and values of local contexts. 

\bibliographystyle{plainnat}
\bibliography{bibliography,bibliography_additional}

\end{document}